\documentclass[aps,prd,twocolumn,superscriptaddress,nofootinbib,10pt]{revtex4-1}

\usepackage[utf8]{inputenc}
\usepackage{graphicx,amsmath,amsfonts,amssymb,aas_macros,slashed}
\usepackage{datetime}
\usepackage{numprint}
\usepackage{enumitem}
\usepackage{graphicx,color,amsmath}
\usepackage{hyperref}

\usepackage{rotating}
\allowdisplaybreaks

\newcommand{\cm}{\ensuremath{\mathrm{cm}}}

\newcommand{\gff}{g_\text{ff}}

\newcommand{\eg}{\textit{e.g.}}

\usepackage{pgfplots}
\usepackage{bm}
\usepackage{mathtools}

\makeatletter
\newcommand{\thickhline}{%
    \noalign {\ifnum 0=`}\fi \hrule height 1pt
    \futurelet \reserved@a \@xhline
}
\newcolumntype{"}{@{\hskip\tabcolsep\vrule width 1pt\hskip\tabcolsep}}
\makeatother

\begin{document}

\title{Is there a radio excess from the decoupling of pre-recombination bremsstrahlung?}

\author{Josef Pradler}
\email{josef.pradler@oeaw.ac.at}
\affiliation{Institute of High Energy Physics, Austrian Academy of Sciences, Georg-Coch-Platz 2, 1010 Vienna, Austria}
\affiliation{CERN, Theoretical Physics Department, 1211 Geneva 23, Switzerland}

\begin{abstract}
Recently it has been suggested that thermal bremsstrahlung emission, when it decouples prior to recombination, creates an excess over the Planck cosmic microwave background spectrum at sub-GHz frequencies. Remarkable by itself, this would also explain a long-standing unexplained deficit in the predictions of the extragalactic radio background. In this brief note we reiterate that no such non-thermal component can arise by itself when matter and radiation remain kinetically coupled.
\end{abstract}

\maketitle

\section{Introduction}
Can a substantial spectral distortion in the cosmic microwave background (CMB) radiation develop from the decoupling of a thermal process that helped maintain the blackbody form in the first place? It was recently suggested that this may indeed happen by the free-free emission process, Bremsstrahlung, in the pre-recombination early Universe at redshifts of a few thousand~\cite{Balaji:2022msm}. An excess of photons for comoving frequencies of 10~MHZ to 1~GHZ  in the Rayleigh-Jeans tail of the CMB is found, surpassing the Planckian occupation number. This would not only amount to a remarkable addition to our understanding of CMB formation, but may also explain a to-date unexplained tension in the extragalactic radio background. When expectations based on faint source counts are confronted with observations, a deficit in the predictions is found~\cite{Protheroe:1996si,Gervasi:2008rr}. For example, at a frequency of $\nu = 200$~MHz the differential Planckian photon number density is $0.01\ \cm^{-3}\ {\rm GHz}^{-1}$. 
Filling the gap and explaining the radio data requires an excess $\Delta n/n_{\rm Planck}|_{200~{\rm MHz}} = 100\%$, where it is surmised that the ``non-Planckian'' build-up of photons from of the free-free process happens from $z=2150$ until recombination. Finally, an extra abundance of photons in the purported frequency range would also amplify~\cite{Fraser:2018acy,Pospelov:2018kdh} the predictions  for the much sought-after cosmological 21~cm signal~\cite{Pritchard:2011xb}, especially given the recently claimed observations~\cite{Bowman:2018yin}.

Evidently, the claim of a {\it guaranteed} non-thermal $O(1)$ distortion of the CMB warrants scrutiny. It is the purpose of this short note to recall standard arguments from kinetic theory and to clarify that no such build-up  from thermal initial conditions is possible.

\section{Bootstrap spectral distortion?}

We now follow through a chain of standard arguments that shows that a {\it large} spectral distortion cannot be seeded by thermal pre-recombination bremsstrahlung alone. For this purpose we ignore any effects that may originate from the mismatch of photon and electron temperatures. Compton scattering ensures their equivalence, $T_e=T$, to high precision until and during recombination.

Denote the isotropic and homogeneous energy-differential number density of photons  by $dn/d\omega$. The associated occupation number is denoted by $f(\omega)$ with $dn/d\omega = (\omega/\pi)^2 f(\omega)$. If a change in $n$ is due to the emission and absorption of single quanta in process $A$, we may write, 
\begin{align}
\label{rateeqn}
   \left. \frac{\partial}{\partial t} \frac{dn}{d\omega}\right|_{A} = \frac{d\Gamma}{d\omega dV} \left( 1 + f\right)  - \frac{dn}{d\omega} \Gamma_{\rm abs}.
\end{align}
Here, ${d\Gamma}/{d\omega dV} $ is the spontaneous emission rate per volume and energy, corrected for stimulated emission by the factor $(1+f)$. The second term on the right hand side accounts for absorption with total rate $\Gamma_{\rm abs}(\omega)$.
In thermal equilibrium, Eq.~\eqref{rateeqn} must vanish identically by the principle of detailed balancing. Hence, plugging in the equilibrium distribution $f_{T} = [\exp{(\omega/T)}-1]^{-1}$ yields the relation between forward and inverse process, 
\begin{align}
\Gamma_{\rm abs} = \frac{\pi^2}{\omega^2} e^{\omega/T} \frac{d\Gamma}{d\omega dV}  .
\end{align}

In an expanding FRW Universe, the Boltzmann equation for the evolution of the photon occupation number $f(\omega)$ under the influence of $A$ is given by 
\begin{align}
\label{boltzmanneqn}
    \frac{\partial f}{\partial t} - H \omega  \frac{\partial f}{\partial \omega} =  \left. \frac{\partial f}{\partial t}\right|_{A} .
\end{align}
Introducing the dimensionless and comoving variable $x = \omega / T$ scales out the expansion term on the left-hand-side when we are to consider the evolution of $f(x)$. A photon number changing process such as double Compton scattering or bremsstrahlung, together with Compton scattering, will bring the photon spectrum to a blackbody form when the rates are faster than the Hubble rate; for an explicit demonstration, see Figs.~1 and 2 of~\cite{1981ApJ...244..392L}.

We are interested in a {\it deviation} from the Planck spectrum that may develop as the Universe cools down. To this end, we write 
\begin{align}
f(x) = f_T(x) + \Delta f(x)  
\end{align}
where $\Delta f>0$ would mean an excess of photons compared to a blackbody. 
By construction, terms with $f_T$ drop out of~\eqref{boltzmanneqn}, and one  is left with an equation that describes the departure from the blackbody spectrum, 
\begin{align}
\label{masterboltz}
    \frac{\partial \Delta f}{\partial t} =  \frac{\pi^2}{x^2 T^3} \frac{d\Gamma}{dx dV}   \Delta f \left( 1 - e^x  \right) .
\end{align}

We may now focus on the pre-recombination era at redshift $z \lesssim 10^4 $ where it is claimed that a large deviation from bremsstrahlung decoupling may be imprinted onto the CMB. This epoch is commonly referred to $y$-distortion era where Compton and double Compton scattering have become inefficient in changing photon number and momenta, respectively~\cite{Chluba:2011hw}. The remaining channel for photon emission is then non-relativistic Bremsstrahlung. Here, the
dominant contribution is dipole emission in the  collision of electrons with protons,  $e p \to ep \gamma$ with an $O(1)$ correction from helium~\cite{1961ApJS....6..167K,vanHoof:2014bha,Chluba:2019ser}; the quadrupole process from electron scattering, $ee\to ee\gamma$, is  suppressed~\cite{Pradler:2020znn,Pradler:2021ppc}. The emission rate in a Maxwellian plasma where matter is in equilibrium with radiation with common temperature $T$ is given by
\begin{align}
\label{dipolerate}
 \frac{d\Gamma}{dx dV} = \frac{16}{3} \sqrt{\frac{2\pi}{3}} \frac{\alpha^3 n_e n_p}{m_e^{3/2} T^{1/2} x} \langle\gff\rangle .
\end{align}
Here, $\alpha$ is the fine-structure constant, $m_e$ is the electron mass, $n_{e}\approx n_p$ are the (approximate) electron and proton number densities and $\langle \gff \rangle$ is the thermally averaged Gaunt factor; see~\cite{Pradler:2021lus} for an expression for $\gff$ and definition of $\langle g_{\rm ff}\rangle$ that covers all non-relativistic kinematic regimes.%
\footnote{The thermally averaged Gaunt factor $\langle \gff\rangle (x) $ is often written in the form $e^{-x} \gff(x)$ where the exponential is then part of the thermally averaged Kramers emissivity.} We note that the emission process is not in the Born regime and  results exact to all orders in the Coulomb interaction of the colliding particle pair must be used~\cite{Sommerfeld:1935ab}; for quadrupole emission cf.~\cite{Pradler:2020znn,Pradler:2021ppc}. To see this, one evaluates the Sommerfeld parameter for a typical relative initial velocity $\eta = Z^2 \alpha/v \simeq Z^2\alpha \sqrt{m_e/T} = 6(24)$ to $10(40)$ for redshifts $z=3000$ to $1000$ for protons (fully ionized helium).

The solution of~\eqref{masterboltz} is readily obtained and given by
\begin{align}
\label{solution}
    \Delta f (x,z=0) =  \Delta f (x,z_{\rm high}) e^{-\tau_{\rm ff}(z_{\rm high})}
\end{align}
where $z_{\rm high}$ is some pre-recombination redshift and $\tau_{\rm ff}$ is the optical depth due to bremsstrahlung,
\begin{align}
    \tau_{\rm ff}(z) = \int_0^z dz\, \frac{1}{(1+z)H} \frac{\pi^2}{x^2 T^3} \frac{d\Gamma}{dx dV}  \left( 1 - e^x  \right) 
\end{align}
which is a strictly positive quantity. For example, a contribution at $\nu=200$~MHz (1~GHz) requires emission with $x=0.02$ $(x=0.1)$. Indeed, the associated optical depth is small $\tau_{ff}|_{z=2000} = 0.003$ ($8\times 10^{-5}$) and the Universe is already transparent for frequencies $\nu\gtrsim 10~$MHz. 

However, as is evident from~\eqref{solution}, a thermal plasma in equilibrium with matter cannot, by itself, develop a deviation from the Planck spectrum. Free-free emission in a medium where photons and electrons share the same temperature will only {\rm reduce} any preexisting spectral distortion. 
Indeed, only if some non-thermal deviation in form of $\Delta f (x,z_{\rm high})\neq 0$ at  initial redshift $z_{\rm high}$ was already present---not produced by thermal bremsstrahlung---may such distortion survive until today. Prerequisite for it is that $\tau_{\rm ff}\ll 1$. 

The statements made here are not new, but appear in various and explicit forms in a broad and long history of studies on thermalization and spectral distortions of the CMB, see, \eg,~\cite{1969Ap&SS...4..301Z,1981ApJ...244..392L,Chluba:2011hw,Chluba:2015hma} and references therein. The spectral distortions that are predicted from the standard recombination cosmic history, \eg, induced by the late-time mismatch of electron and photon temperature, are, compared to the claim made in~\cite{Balaji:2022msm}, minute.

\paragraph*{Acknowledgements} 
The author thanks C.~Boehm, X.~Chu and L.~Semmelrock for useful discussions. 

\bibliography{refs}

\end{document}